\documentclass{ws-Preprint} 
\begin{document}
\title{\Large LIGHT CLUSTER PRODUCTION IN CENTRAL SYMMETRIC HEAVY-ION REACTIONS FROM FERMI TO GeV ENERGIES.}
\author{R\'emi Bougault $^{1}$, 
Bernard Borderie $^{2}$,
Abdelouahad Chbihi $^{3}$,
Quentin Fable $^{1}$,
John David Frankland $^{3}$,
Emmanuelle Galichet $^{2,4}$,
Tom Genard $^{3}$,
Di\'ego Gruyer $^{1}$,
Maxime Henri $^{3}$,
Marco La Commara $^{5}$,
Nicolas Le Neindre $^{1}$,
Ivano Lombardo $^{8}$,
Olivier Lopez $^{1}$,
Marian P\^arlog $^{1}$,
Piotr Paw{\l}owski $^{6}$,
Giuseppe Verde $^{7,8}$,
Emmanuel Vient $^{1}$,
Mariano Vigilante $^{5}$\\
(INDRA collaboration)
\vspace{6pt}
}
\address{%
$^{1}$ Normandie Univ, ENSICAEN, UNICAEN, CNRS/IN2P3, LPC Caen, France\\
$^{2}$ Universit\'e Paris-Saclay, CNRS/IN2P3, IJCLab, Orsay, France\\
$^{3}$ Grand Acc\'el\'erateur National d'Ions Lourds (GANIL), CEA/DRF-CNRS/IN2P3, Caen, France\\
$^{4}$ Conservatoire National des Arts et Metiers, Paris, France\\
$^{5}$ Universit\'a "Federico II", Naples, Italy\\
$^{6}$ IFJ PAN, Cracow, Poland\\
$^{7}$ Laboratoire des 2 Infinis Toulouse, Universit\'e de Toulouse, CNRS/IN2P3, UPS, Toulouse, France\\
$^{8}$ Istituto Nazionale di Fisica Nucleare (INFN), Sezione di Catania, Catania, Italy\\
\vspace{6pt}
(corresponding author: bougault@lpccaen.in2p3.fr)
}
%%%%%%%%%%%%%%%%%%
\maketitle
% Contact information of the corresponding author
%\corres{Correspondence: bougault@lpccaen.in2p3.fr}
%%%%%%%%%%%%%%%%%%%%%%%%%%%
\abstracts{\underline{Abstract:}
Correlations and clustering are of great importance in the study of the Nuclear Equation of State. Information on these items/aspects
can be obtained using Heavy-Ion reactions which are described by dynamical theories. We propose a dataset that will be useful for improving the description of light cluster production in transport model approaches. The dataset combines published and new data and is presented in a form that allows direct comparison of the experiment with theoretical predictions. The dataset is ranging in bombarding energy from 32 to 1930 A MeV. In constructing this dataset we put in evidence the existence of a change in the light cluster production mechanism that corresponds to a peak in deuteron production.}
% Keywords
%\keyword{nuclear physics; heavy-ion reactions; nuclear equation of state; %light cluster production; transport models} 
%%%%%%%%%%%%%%%%%%%%%%%%%%%%%%%%%%%%%%%%%%
%%%%%%%%%%%%%%%%%%%%%%%%%%%%%%%%%%%%%%%%%%
\section{Introduction}
Knowledge of the mechanism of fragment and light cluster formation in heavy-ion collisions allows to trace the fundamental properties of nuclear matter~[\cite{Ono2019}].
%%%% 
In nuclear collisions the emission of several fragments from a hot nucleus, multifragmentation, has been observed and the fragment characteristics are well described by statistical concepts. But in multifragmentation models the description of the total observed production rates of light particles does not reach, by far, the accuracy obtained for the heavier fragments.
%%%%
In heavy-ion transport models two main actors are at work, the one-body approach (mean field) and the few body correlation in a medium (clustering). Formation of clusters and fragments are determined by proper treatment of correlations and proper introduction and propagation of fluctuations in dynamical models. It turns out that a problem exists concerning light cluster description in many transport approaches. This problem has been addressed, for example, in~[\cite{Danielewicz1992}] with dynamic production of A$\leq$3 fragments and in~[\cite{Ono2019}] by considering wave-packet splitting and
by taking into account explicitely the cluster correlations.
In all cases, the calculations have shown that taking into account the clustering is important since many observables depend on it.
\par
The aim of the present article is to provide experimental data on cluster production 
to which the model calculations can be directly compared in order to improve our understanding of the Nuclear Equation of State.
It is also useful to recall that light nuclear clusters play an important role in the warm
and low-density nuclear matter that can be found in core-collapse supernovae and neutron star mergers~[\cite{Pais2020}] and therefore their production mechanism is also important for astrophysics.
%%%%%%%%%%%%%%%%%%%%%%%%%%%%%%%%%%%%%%%%%%
\section{Materials and Methods}
The $4\pi$ multi-detector INDRA~[\cite{Pouthas1995}] 
was used to study four nuclear reactions with beams of $^{58}$Ni, 
accelerated by the GANIL cyclotrons (Caen) to 32, 52, 64, 74 MeV/nucleon, and thin (179 $\mu$g/cm$^{2}$) target of $^{nat}$Ni,
and two reactions with beams of $^{197}$Au, accelerated by
the heavy-ion synchrotron SIS at GSI (Darmstadt) to 40, 60 MeV/nucleon, and 2000 $\mu$g/cm$^{2}$ target of $^{nat}$Au.
Higher beam energy experiments were performed during these experimental campaigns, but they were 
excluded from the present analysis because of the limited stopping power of the experimental INDRA apparatus
to high energetic light charged particles produced in the forward direction. 
\par
INDRA is a charged product multidetector,
composed of 336 detection cells arranged in 17 rings centered
on the beam axis and covering 90\% of the solid angle. 
The first ring (2$^{o}$ to 3$^{o}$), made of 12 phoswich telescopes, was not used in the Ni+Ni analysis due to a malfunction during the experiment. For Au+Au experimental campaign, the phoswich telescopes (2$^{o}$ to 3$^{o}$) were replaced by 12 telescopes each one composed of 300 $\mu$m
silicon wafer (Si) and a CsI(Tl) scintillator (14 cm thick). Rings 2 to 9 (3$^{o}$ to 45$^{o}$) are composed of 12 or 24
three-member detection telescopes : a 5 cm thick ionization
chamber (50 mbar); a 300 $\mu$m silicon
wafer; and a CsI(Tl) scintillator (14 to 10 cm thick) coupled
to a photomultiplier tube. 
Rings 10 to 17 (45$^{o}$ to 176$^{o}$) are composed of 24, 16 or 8 two-member telescopes:
an ionization chamber (5 cm thick, 30 mbar) and a CsI(Tl) scintillator of 8, 6
or 5 cm thickness.
INDRA can identify in charge fragments from Hydrogen to Uranium and in mass light fragments (Z$\leq$ 4) with low thresholds.
Recorded event functionality was activated under a triggering factor based on a minimum number of fired 
telescopes (N$^{min}$) over the detector acceptance (90\% of $4\pi$). 
During the Ni+Ni experiments, N$^{min}$ was set to 4 while during the Au+Au experiments, N$^{min}$ was 3.
\par
The goal of the present article is to extend to low bombarding energies some of the results presented
in an article published by the FOPI collaboration~[\cite{Reisdorf2010}]: the yields of light clusters contained in appendix A.
The FOPI article ''Systematics of central heavy ion collisions in the 1$A$ GeV regime'' presents data
concerning 25 system-energies from 90$A$ MeV to 1.93$A$ GeV bombarding energies. Because for a given bombarding energy cluster yields depend on projectile and target isotopic composition~[\cite{Bougault2018}], we will use the systems listed in Table~\ref{tab0} to perform the comparison.
\begin{table}[!h] 
\centering
\begin{tabular}{|c|c|c|c|c|}
\hline
{\textbf{Projectile and Target}}	& \raisebox{-0.5ex}{\textbf{$^{58}$Ni+$^{nat}$Ni}}	& 
\raisebox{-0.5ex}{\textbf{$^{58}$Ni+$^{58}$Ni}}& \raisebox{-0.5ex}{\textbf{$^{40}$Ca+$^{40}$Ca}}& 
\raisebox{-0.5ex}{\textbf{$^{197}$Au+$^{nat}$Au}}\\
\hline
{Data set}	& {INDRA} & {FOPI} & {FOPI} & {INDRA and FOPI}\\
\hline
{N/Z} & {1.084} & {1.071}  & {1.000} & {1.493} \\
\hline
\end{tabular}
\caption{List of the systems (projectile+target) retained for the present analysis. The data set refers to the used experimental apparatus. The last row presents the neutron to proton ratio of the total combined system.}
\label{tab0}
\end{table}
\par
The small N/Z difference between Ni+Ni and Ca+Ca indicates that these two data sets can be aggregated when using independent system size variables. 
%%%%%%%%%%%%%%%%%%%%%%%%%%%%%%%%%%%%%%%%%%
\section{Results}
\subsection{Central event selection and cluster mean multiplicities}
The FOPI data, yield tables presented in appendix A of~[\cite{Reisdorf2010}], correspond to a centrality selection with an estimated upper limit of reduced impact parameter (b$_{0}$) of 0.15. Furthermore a carefull work of interpolations and extrapolations has been performed so as to present event topology in full 4$\pi$ coverage from detected events. This last point explains the rather large multiplicity uncertainties presented in appendix A.
\par
For INDRA data the total transverse energy of detected light charged particles (lcp, Z=1 and 2), $(\Sigma E_{t})^{lcp}$,  is chosen as an impact parameter selector, and, as for the FOPI data, centrality selections were defined using the sharp cut-off approximation of~[\cite{Cavata1990}]  in order to have estimated reduced impact parameters for central events 
b$_{0}$<0.15. For INDRA data, the lcp yields are calculated using only the forward part of the center of mass (hereinafter called c.m.). The multi-detector, for these reactions, possesses better detection performances in the forward c.m. half hemisphere as compared to the backward part. So, if necessary, the total detected yields can be estimated by doubling the values.
\par
In Tables~\ref{tab1} and~\ref{tab2} are presented, for each system, the  $(\Sigma E_{t})^{lcp}$ threshold value used for central event selection and the mean value of cluster multiplicities detected in the forward c.m. part. The associated uncertainty values correspond to the error on the mean value determination (standard error on the mean). 
% TABLES 1 et 2  %%%%%%%%%%%%%%%%%%%%%%%%%%%%%%%%%%%%%%%%%%%%%%%%%%%%%%
\begin{table}[!h] 
\centering
\begin{tabular}{|c|c|c|c|c|}
\hline
\raisebox{-0.5ex}{\textbf{$^{58}$Ni+$^{nat}$Ni (b$_{0}~<$ 0.15)}}	& {\textbf{32 A MeV}}	& {\textbf{52 A MeV}} & {\textbf{64 A MeV}} & {\textbf{74 A MeV}}\\
\hline
\raisebox{-0.5ex}{$(\Sigma E_{t})^{lcp}~>$}	& {225 MeV} &	 {350 MeV} & {425 MeV} & {500 MeV}\\
\hline
{$(M_{^{1}H})_{forward~c.m.}$} & {3.000$\pm$0.003} & {3.582$\pm$0.003} & {3.866$\pm$0.003}  & {3.972$\pm$0.003} \\
\hline
{$(M_{^{2}H})_{forward~c.m.}$} & {1.049$\pm$0.002} & {1.583$\pm$0.002} & {1.823$\pm$0.002}  & {1.948$\pm$0.002} \\
\hline
{$(M_{^{3}H})_{forward~c.m.}$} & {0.447$\pm$0.001} & {0.753$\pm$0.002} & {0.958$\pm$0.002}  & {1.090$\pm$0.002} \\
\hline  
{$(M_{^{3}He})_{forward~c.m.}$} & {0.340$\pm$0.001} & {0.570$\pm$0.001} & {0.695$\pm$0.001}  & {0.793$\pm$0.002} \\
\hline      
{$(M_{^{4}He})_{forward~c.m.}$} & {3.116$\pm$0.003} & {3.491$\pm$0.004} & {3.604$\pm$0.003}  & {3.557$\pm$0.003} \\ 
\hline        
\end{tabular}
\caption{Ni+Ni central events (INDRA), four bombarding energies: threshold values of the total transverse energy of detected light charged particles which correspond to a reduced impact parameter of 0.15. Mean values of cluster multiplicities detected in the forward c.m. part and their associated uncertainties.} 
\label{tab1}
\end{table}
\begin{table}[!h]
\centering 
\begin{tabular}{|c|c|c|}
\hline
\raisebox{-0.5ex}{\textbf{$^{197}$Au+$^{nat}$Au (b$_{0}~<$ 0.15)}}	& {\textbf{40 A MeV}}	& {\textbf{60 A MeV}}\\
\hline
\raisebox{-0.5ex}{$(\Sigma E_{t})^{lcp}~>$}	& {625 MeV} &	 {1050 MeV}\\
\hline
{$(M_{^{1}H})_{forward~c.m.}$}	& {4.016$\pm$0.005} & {5.185$\pm$0.003}\\
\hline
{$(M_{^{2}H})_{forward~c.m.}$}	& {2.484$\pm$0.004} & {3.412$\pm$0.002}\\
\hline
{$(M_{^{3}H})_{forward~c.m.}$}	& {2.107$\pm$0.004} & {2.837$\pm$0.002}\\
\hline
{$(M_{^{3}He})_{forward~c.m.}$}	& {0.731$\pm$0.002} & {1.179$\pm$0.002}\\
\hline
{$(M_{^{4}He})_{forward~c.m.}$}	& {5.843$\pm$0.002} & {6.816$\pm$0.003}\\
\hline
\end{tabular}
\caption{Au+Au central events (INDRA), two bombarding energies: threshold values of the total transverse energy of detected light charged particles which correspond to a reduced impact parameter of 0.15. Mean values of cluster multiplicities detected in the forward c.m. part and their associated uncertanties.} 
\label{tab2}
\end{table}
%%%%%%%%%%%%%%%%%%%%%%%%%%%%%%%%%%%%%%%%%%%%%%%%%%%%%%%%%%%%%%%%%%%%%%%
\subsection{Cluster production: abundance ratios}
As mentioned previously, the FOPI data is extrapolated to full 4$\pi$ coverage and therefore it is impossible to compare directly mean multiplicity values presented in~[\cite{Reisdorf2010}] to INDRA data presented in Tables~\ref{tab1} and~\ref{tab2}.
The FOPI data is extrapolated because the detection efficiency varies according to the bombarding energy. 
If this variation is not corrected the different bombarding energy data can hardly be compared. This variation according to bombarding energy is also true for INDRA data. An example of this defect can be seen by comparing the INDRA multiplicity of $^{4}$He for the 64 and 74 A MeV Ni+Ni systems (Table~\ref{tab1}): the multiplicity is decreasing as the bombarding energy is increasing simply because of the absence of a detector below three degrees in the laboratory reference frame.   
\par
We have chosen not to extrapolate the INDRA data using the cluster abundance ratios~[\cite{Gutbrod1989}].	One way round the varying detection efficiency is to compare cluster mean multiplicities relative to proton mean multiplicity (hereafter called cluster abundance ratios). Doing so, for each system we can hope that the detection efficiency variation is largely canceled out since it is present in both the numerator and the denominator of the cluster abundance ratio. 
The polar angular area not covered by INDRA is from 0 to 3 degrees for Ni+Ni and 0 to 2 degrees for Au+Au. This is a source of systematic errors in addition to the statistical errors presented in Table~\ref{tab1} and~\ref{tab2} to calculate the total abundance ratios uncertainties if those ratios were calculated with extrapolated 4$\pi$ multiplicities. In the present case, we do not take this into account because the values of the INDRA ratios are calculated with the measured multiplicities.
The use of abundance ratios, or even other ratios, allows also to compare the FOPI full 4$\pi$ coverage data set to the INDRA forward c.m. detected data set.
The other advantage lies in the fact that ratios remove also trivial size dependency~[\cite{Reisdorf2010}] and therefore we will be authorized to aggregate Ni+Ni with Ca+Ca results since these systems have almost the same global neutron to proton ratio (which we will hereafter call isospin by abuse of language).
\par
\begin{figure}[!ht]
\centering
\resizebox{0.75\textwidth}{!}{%
   \includegraphics{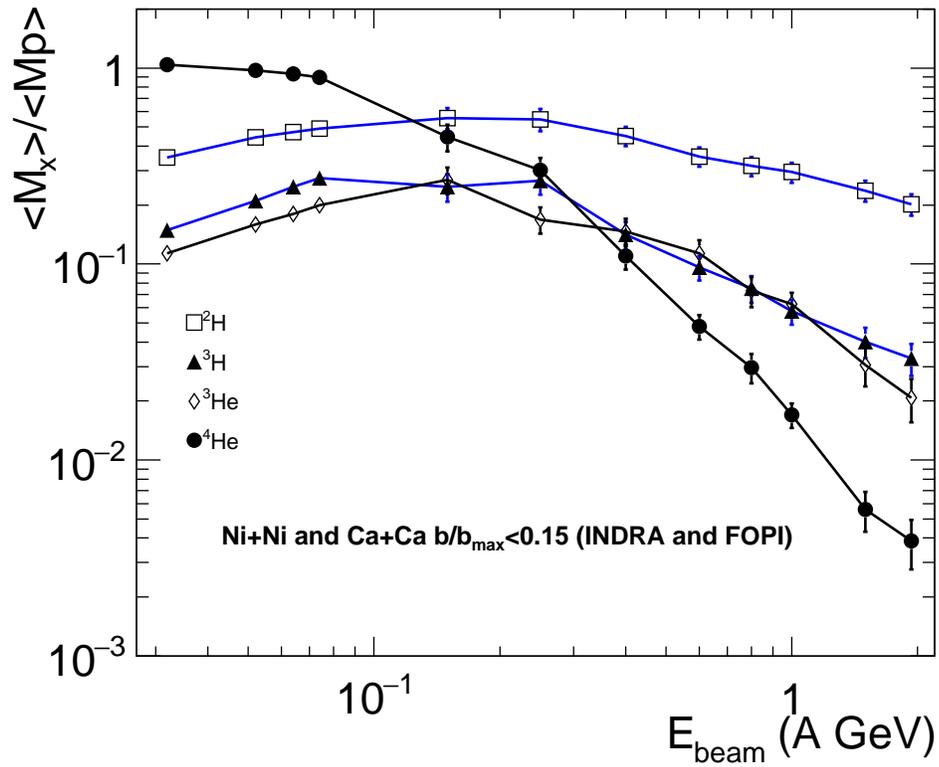}
   } 
\caption{Light cluster abundance ratios as a function of bombarding energy for Ni+Ni and Ca+Ca systems. Lines to guide the eye. Line colours are to differentiate them.}
\label{fig1}
\end{figure}
\begin{figure}[!ht]
\centering
\resizebox{0.75\textwidth}{!}{%
   \includegraphics{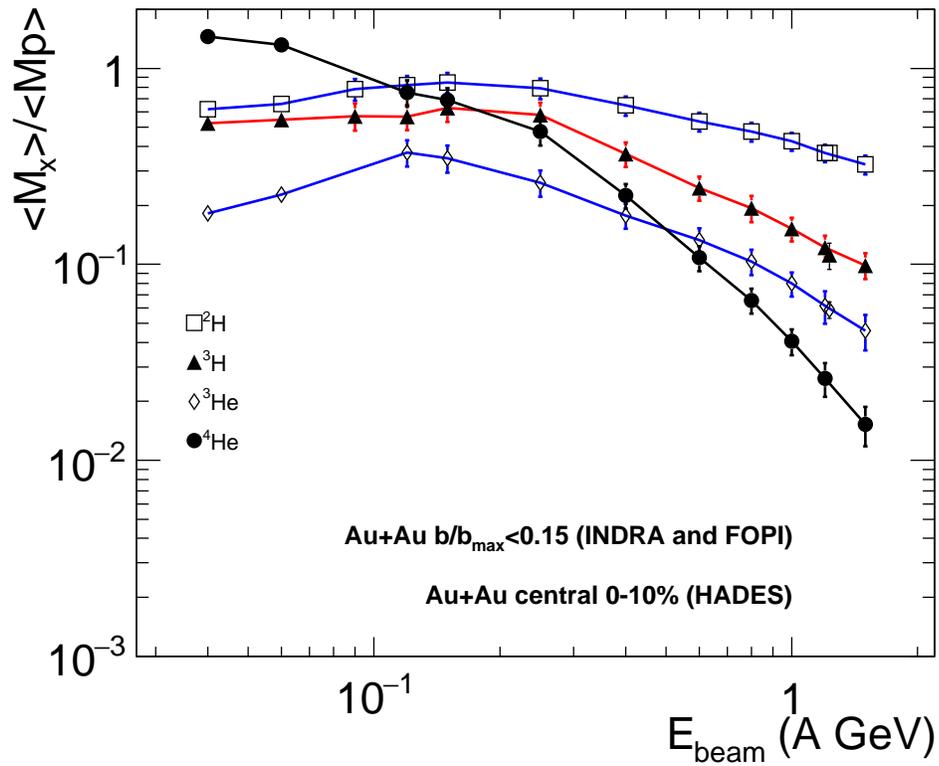}
   } 
\caption{Light cluster abundance ratios as a function of bombarding energy for Au+Au. Lines to guide the eye. Line colours are to differentiate them.}
\label{fig2}
\end{figure}

$^{2}$H, $^{3}$H, $^{3}$He and $^{4}$He abundance ratios are presented in Figure~\ref{fig1} for Ni+Ni and Ca+Ca systems and in Figure~\ref{fig2} for Au+Au.
\par
In Figure~\ref{fig1}, bombarding energy is ranging from 32 to 1930 A MeV. The FOPI Ca+Ca data starts at 400 A MeV. Below 400 A MeV the data are Ni+Ni systems, the first four values concern INDRA data set. 
The absence of discontinuity between values concerning Ni+Ni and Ca+Ca systems confirms the fact that the use of ratios eliminates trivial size effects.
\par
In Figure~\ref{fig2}, bombarding energy is ranging from 40 to 1500 A MeV. The FOPI data starts at 90 A MeV for which $^{3}$He and $^{4}$He multiplicity values are not available in the published article. 
HADES data for $^{2}$H, $^{3}$H and $^{3}$He~[\cite{ref-url1}] at 1230 A MeV bombarding energy are also included in Figure~\ref{fig2}. The HADES data correspond to the 10\% most central events~[\cite{Adamczewski-Musch2018}]. 
The multiplicity values of HADES are not extrapolated over 4$\pi$ and by comparing them to adjacent FOPI multiplicity values (1200 A MeV) [\cite{Reisdorf2010}] we find that they are very different (for example the FOPI proton multiplicity is 99.3 as compared to 77.6 for the HADES proton multiplicity). 
The figure shows, however, that the abundance ratio values are very close: HADES and FOPI results are compatible. We therefore see that the use of multiplicity ratios cancels out the detection inefficiencies.
\par
Apart from the fact that the absolute values of cluster abundance ratios are not the same for Au+Au and Ni+Ni/Ca+Ca systems because of different isospin values, we can still observe some common trends between the two Figures. 
As the bombarding energy is increasing, we note a dramatic decrease of the $^{4}$He abundance ratio. Simultaneously the other cluster abundance ratios are increasing to a maximum value which is reached at about 150 A MeV; this is particularly true for $^{2}$H. Above 200 A MeV all cluster abundance ratios are decreasing with increasing bombarding energy. The $^{4}$He abundance ratio decreases even more rapidly as compared to the others.    
\par
Different mechanisms of cluster production are at work in central heavy ion collisions. In particular at moderate bombarding energy, part of this production is due to secondary de-excitation from fragments produced with internal excitation energy. Nevertheless, the common trends observed in Figures~\ref{fig1} and~\ref{fig2} are significant. 
%%%%%%%%%%%%%%%%%%%%%%%%%%%%%%%%%% 
\subsection{Cluster production: selected multiplicity ratios}
\begin{figure}[!ht]
\centering
\resizebox{0.75\textwidth}{!}{%
   \includegraphics{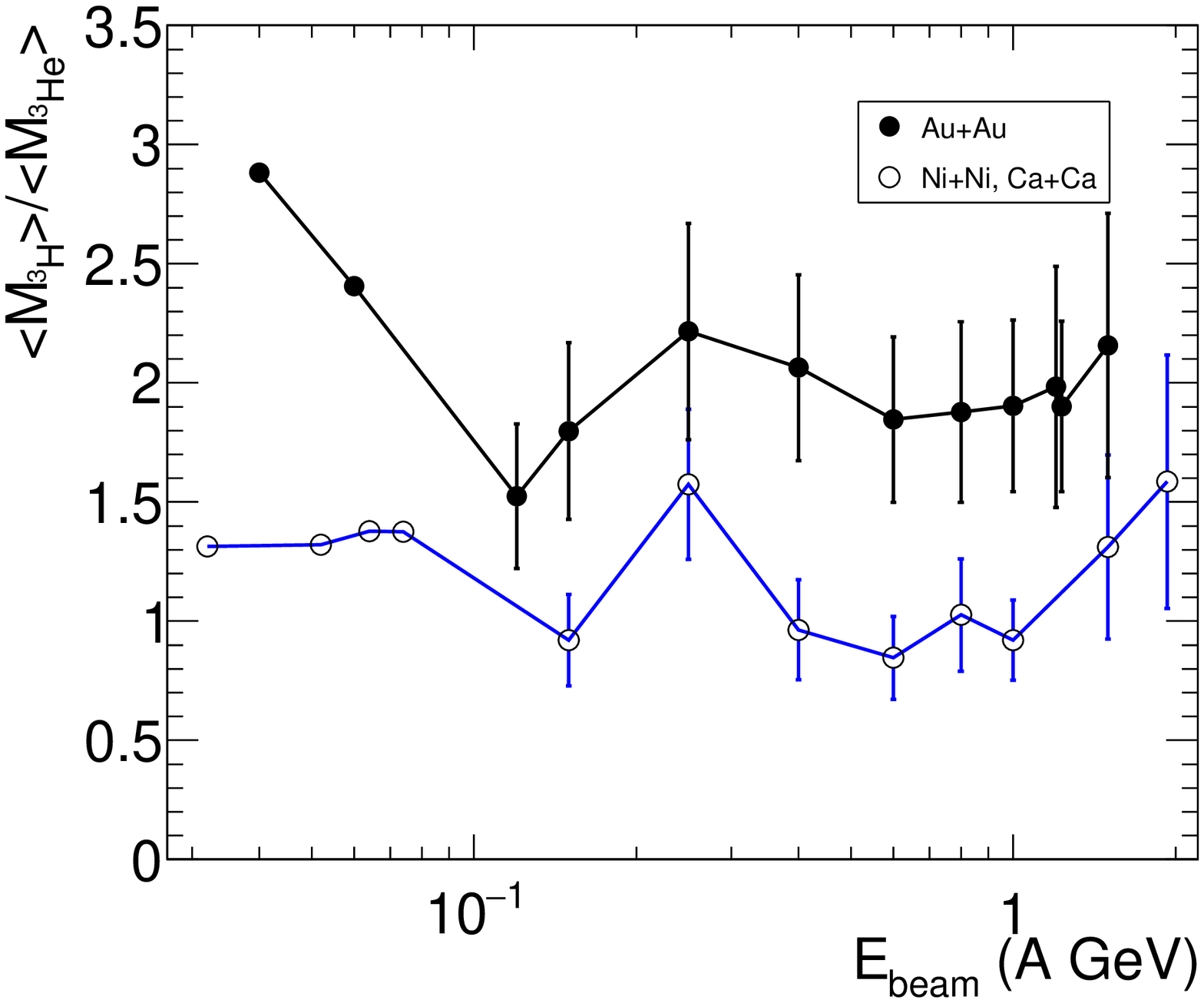}
   }
\caption{Ratio of $^{3}$H to $^{3}$He mean multiplicities as a function of bombarding energy for Au+Au and Ni+Ni/Ca+Ca systems. Lines to guide the eye. Black line corresponds to Au+Au, blue line to Ni+Ni/Ca+Ca.}
\label{fig3}
\end{figure}

The $^{3}$H and $^{3}$He productions are strongly isospin dependent~[\cite{Bougault2018}], this is visible also comparing Figures~\ref{fig1} and~\ref{fig2}. 
We have plotted in Figure~\ref{fig3} the ratio of mean multiplicities of the two species for Au+Au and Ni+Ni/Ca+Ca systems. 
It is seen that the use of the ratio of mean multiplicities of $^{3}$H and $^{3}$He does not cancel the isospin dependence. The ratio is bigger for the Au+Au system whose isospin value is the greatest (see Table~\ref{tab0}).
For an ideal gas scenario this ratio is related to free neutron to free proton ratio~[\cite{Albergo1985}].
This would imply a simultaneous emission of all species thus ignoring a possible existence of a temporality in the lcp emission process~\cite{Xi1998,Verde2002,Chen2003}. 
For example, it has been shown experimentally that the characteristics of $^{3}$He production reflect on average a rapid emission~[\cite{Bougault2018}] at Fermi energies. 
It follows that, 
this does not justify using this ratio to measure the ratio of free neutrons to free protons, which at best it reflects over the full energy range studied here.
So this ratio will not be used here to measure a characteristic of a hypothetic lcp emission source but rather to give us a reference to compare the differences between other observables that will be studied next.
\par
\begin{figure}[!ht]
\centering
\resizebox{0.75\textwidth}{!}{%
   \includegraphics{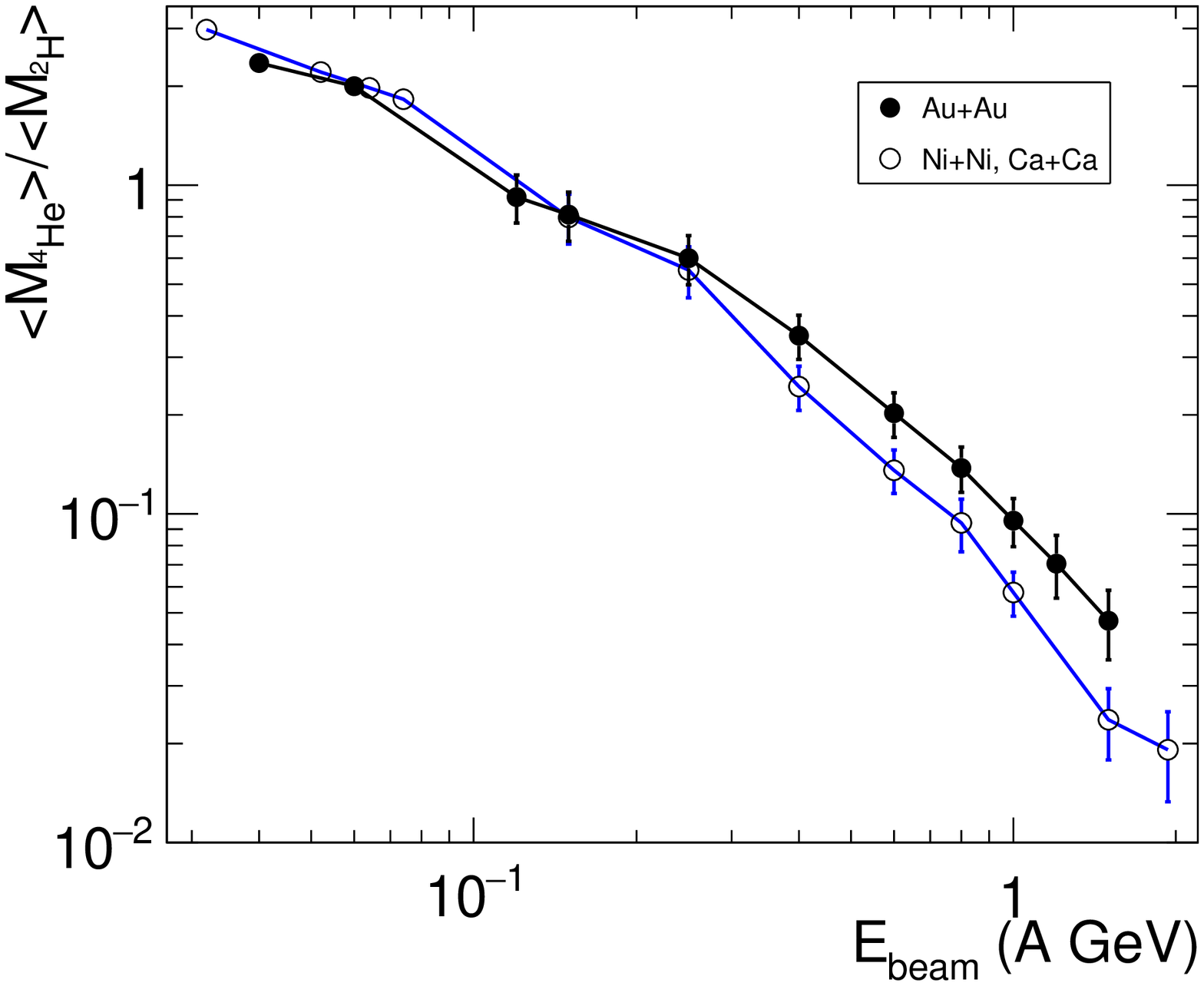}
   } 
\caption{Ratio of $^{4}$He to $^{2}$H mean multiplicities as a function of bombarding energy for Au+Au and Ni+Ni/Ca+Ca systems. Lines to guide the eye. Black line corresponds to Au+Au, blue line to Ni+Ni/Ca+Ca.}
\label{fig4}
\end{figure}

The $^{4}$He to $^{2}$H mean multiplicity ratio is now examined. This ratio is presented in Figure~\ref{fig4}.  
For all systems the ratio is decreasing with increasing bombarding energy. This is reflecting the dramatic decrease of $^{4}$He abundance ratio.
The ratio $^{4}$He/$^{2}$H is almost system independent from few tens A MeV up to about 150 A MeV, the value for which the $^{2}$H abundance ratio reaches its maximum (Figures~\ref{fig1} and~\ref{fig2}). Then increasing the bombarding energy the two curves diverge to reach a relative difference comparable to the one observed for the $^{3}$H over $^{3}$He ratio (Figure~\ref{fig3}). For low bombarding energies, $^{4}$He can be considered as two $^{2}$H and therefore the ratio is almost independent of the system. At higher energies, this independence fades and the heavier particle is less likely to be produced when the system is lighter. This indicates a change of light cluster mean production around few hundred A MeV. 
\par
\begin{figure}[!ht]
\centering
\resizebox{0.75\textwidth}{!}{%
   \includegraphics{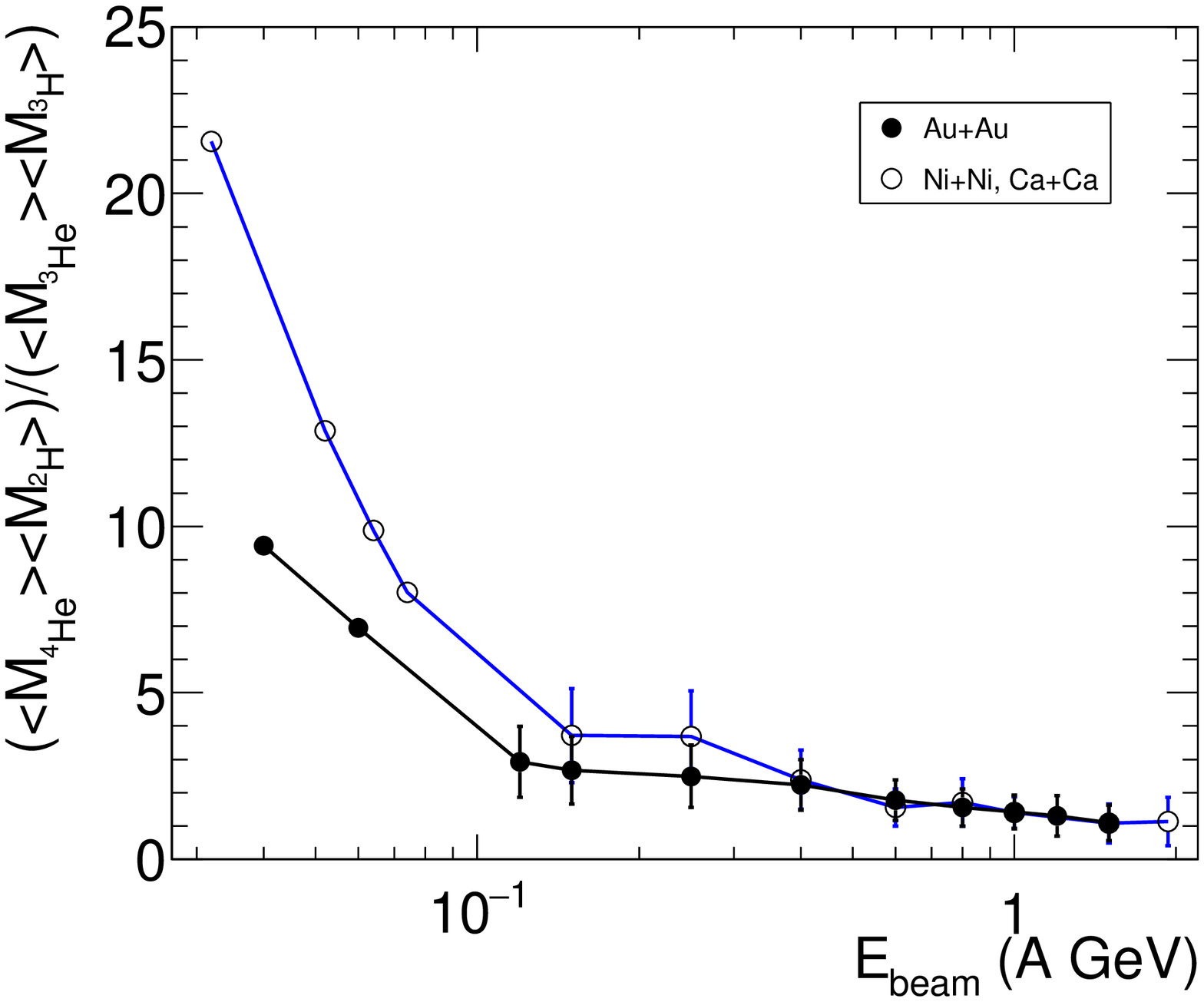}
   } 
\caption{Combined ratio using $^{2}$H, $^{3}$H, $^{3}$He and $^{4}$He mean multiplicities as a function of bombarding energy for Au+Au and Ni+Ni/Ca+Ca systems. Lines to guide the eye. Black line corresponds to Au+Au, blue line to Ni+Ni/Ca+Ca.}
\label{fig5}
\end{figure}

A combined ratio using $^{2}$H, $^{3}$H, $^{3}$He and $^{4}$He mean multiplicities will now be examined. It is presented in Figure~\ref{fig5} as a function of bombarding energy. The ratio with $^{4}$He and $^{2}$H mean multiplicities in the numerator and $^{3}$He and $^{3}$H mean multiplicities in the denominator is directly connected to the temperature of an ideal gas~[\cite{Albergo1985}]. It should be isospin and mass independent in this scenario.
From Figure~\ref{fig5}, according to bombarding energy, it can be seen that: (i) from few tens of A MeV the ratio depends on the system characteristics, (ii) there is a change of slope around 150 A MeV, (iii) from 400 A MeV and above the ratio is system independent. This independence for high bombarding energy is remarkable since, from Figures~\ref{fig3} and~\ref{fig4}, it has been noted a system dependency for $^{3}$H/$^{3}$He and $^{4}$He/$^{2}$H. All the mean multiplicity differences counterbalance from 400 A MeV. 
This could be an indication that on average all detected particles are emitted simultaneously from 400 A MeV onwards, knowing that for lower energies the average values contain the imprint of different processes. This modification starts to be observed around 150 A MeV by a change in the slope of the observable, the bombarding energy value for which the production of $^{2}$H reaches a maximum.
%%%%%%%%%%%%%%%%%%%%%%%%%%%%%%%%%%%%%%%%%%
\section{Discussion}
The aim of the present article is to provide experimental data on cluster production in central collisions 
to which the model calculations can be directly compared.
For this purpose the FOPI data of~[\cite{Reisdorf2010}] were extended to lower bombarding energies, using data obtained with INDRA multidetector both at GANIL (Caen) and GSI (Darmstadt). The results are presented in the form of ratios in order to be less dependent of the experimental set-up and different bombarding energies. The ratios are light cluster abundance ratios, i.e. mean cluster multiplicity over mean proton multiplicity. Also are presented the following ratios: $^{3}$H/$^{3}$He, $^{4}$He/$^{2}$H and $^{4}$He$^{2}$H/$^{3}$He$^{3}$H ($^{A}$X here represents the mean value of cluster multiplicity $^{A}$X). The extension towards low bombarding energies allows to highlight common features concerning cluster production for Au+Au and combined Ni+Ni/Ca+Ca systems, different systems with regard to total mass and total neutron to proton ratio. Increasing the bombarding energy for all systems:
\begin{itemize}
\item There is a dramatic decrease of $^{4}$He abundance ratio; 
\item There exists a clear maximum for $^{2}$H abundance ratio located around 150 A MeV;
\item There exists also a maximum around 150 A MeV for $^{3}$H and $^{3}$He abundance ratios but this maximum is less pronounced.
\end{itemize}
Looking at $^{4}$He/$^{2}$H and $^{4}$He$^{2}$H/$^{3}$He$^{3}$H ratios, it appears that there exists a change of the mechanism of cluster production for that maximum value of 150 A MeV. For low bombarding energies several mechanisms compete, direct production and secondary decay of excited fragments, whereas for high bombarding energies it seems that on average the cluster production is more in line with a common temporality. 
\par
The presented results concern static observables (multiplicity) and they should be seen in conjunction with the results obtained on nuclear stopping in central events~[\cite{Andronic2006,Reisdorf2010,Henri2020}] concerning dynamical observables. Cluster production modelling needs to take both aspects into account. 
%%%%%%%%%%%%%%%%%%%%%%%%%%%%%%%%%%%%%%%
%%%%%%%%%%%%%%%%%%%%%%%%%%%%%%%%%%%%%%%%%%
%%%%%%%%%%%%%%%%%%%%%%%%%%%%%%%%%%%%%%%%%%
\vspace{6pt}
\section*{Acknowledgements}
We thank the GANIL and GSI staff for
providing us the beams and for the technical support during the experiments. We acknowledge
support from Centre National de la Recherche Scientifique/Institut National de Physique Nucl\'eaire et de Physique des Particules (France) and R\'egion Normandie (France) under RIN/FIDNEOS. 
%=====================================

%%%%%%%%%%%%%%%%%%%%%%%%%%%%%%%%%%%%%%%%%%
%%%%%%%%%%%
%%%%%%%%%%%
%%%%%%%%%%%
%%%%%%%%%%%
%%%%%%%%%%%%%%
%%%%%%%%%%%%
%%%%%%%%%%%%%
%%%%%%%%%%%%%

\begin{thebibliography}{999}
\bibitem{Ono2019} 
{Ono, A. Dynamics of clusters and fragments in heavy-ion collisions. 
Prog. Part. Nucl. Phys.
(2019), {\textbf 105}, 139-179.}
%
\bibitem{Danielewicz1992}
{Danielewicz, P.; et al. Blast of light fragments from central heavy-ion collisions. 
Phys. Rev. C
(1992), {\textbf 46}, 2002-2011.}
%
\bibitem{Pais2020}
{Pais, H.; et al. Low Density In-Medium Effects on Light Clusters from Heavy-Ion Data. 
Phys. Rev. Lett.
(2020), {\textbf 125}, 012701.}
%
\bibitem{Pouthas1995} 
{Pouthas, J.; et al. INDRA, a 4$\pi$ charged product detection array at GANIL. 
Nucl. Instr. and Meth. A
(1995), {\textbf 357}, 418--442.}
%
\bibitem{Reisdorf2010} 
{Reisdorf, W.; et al. Systematics of central heavy ion collisions in the 1~A GeV regime. 
Nucl. Phys. A
(2010), {\textbf 848}, 366--427.}
%
\bibitem{Bougault2018}
{Bougault, R.; et al. Light charged clusters emitted in 32 MeV/nucleon $^{136,124}$Xe+$^{124,112}$Sn reactions: Chemical equilibrium and production of $^{3}$He and $^{6}$He. 
Phys. Rev. C
(2018), {\textbf 97}, 024612.}
%
\bibitem{Cavata1990}
{Cavata, C.; et al. , Determination of the impact parameter in relativistic nucleus-nucleus collisions.
Phys. Rev. C 
(1990), {\textbf 42}, 1760-1763.}
%
\bibitem{Gutbrod1989}
{Gutbrod, H.H.; et al. , Plastic Ball experiments.
Rep. Prog. Phys.
(1989), {\textbf 52}, 1267-1328.}
%
\bibitem{ref-url1}
{ECT* Workshop: Light clusters in nuclei and nuclear matter: Nuclear structure and decay, heavy ion collisions, and astrophysics (2019). Available online: 
https://indico.ectstar.eu/event/52/contributions/1434/
(Szala, M.; for the HADES collaboration, contribution accessed on 15 July 2020).}
%
\bibitem{Adamczewski-Musch2018}
{Adamczewski-Musch, J.; et al., Centrality determination of Au + Au collisions at 1.23A GeV with HADES
Eur. Phys. J. A 
(2018), {\textbf 54}, 85.}
%
\bibitem{Albergo1985}
{Albergo, S.; et al., Temperature and Free-Nucleon Densities of Nuclear Matter Exploding into Light Clusters in Heavy-Ion Collisions.
Nuovo Cimento A 
(1985), {\textbf 89}, 1-28.}
%
\bibitem{Xi1998}
{Xi, H.F.; et al., Dynamical emission and isotope thermometry.
Phys. Rev. C 
(1998), {\textbf 58}, R2636-R2639.}
%
\bibitem{Verde2002}
{Verde, G.; et al. , Imaging sources with fast and slow emission components.
Phys. Rev. C
(2002), {\textbf 58}, 054609.}
%
\bibitem{Chen2003}
{Chen, Lie-Wen; et al. , Light cluster production in intermediate energy heavy-ion collisions induced by neutron-rich nuclei.
Nucl. Phys. A 
(2003), {\textbf 729}, 809-834.}
%
\bibitem{Andronic2006}
{Andronic, A.; et al. , Systematics of stopping and flow in Au+Au collisions.
Eur. Phys. J. A 
(2006), {\textbf 30}, 31-46.}
%
\bibitem{Henri2020}
{Henri, M.; et al. , In-medium effects in central heavy ion collisions at intermediate energies.
Phys. Rev. C 
(2020), {\textbf 101}, 064622.}
\end{thebibliography}
\end{document}